\documentclass[conference]{IEEEtran}

\usepackage{acronym}
\usepackage{amsfonts}
\usepackage[dvips]{graphicx}
\usepackage{times}
\usepackage{cite}
\usepackage{amsmath}
\usepackage{array}
\usepackage{amssymb}
\usepackage{stfloats}
\usepackage{diagbox}
\usepackage{graphicx}
\usepackage{footnote}
\usepackage{amsthm}
\usepackage{booktabs}
\usepackage{array}
\usepackage[ruled,vlined]{algorithm2e}
\usepackage{subeqnarray}
\usepackage{cases}
\usepackage{threeparttable}
\usepackage{color}
\usepackage{epstopdf}
\usepackage{multirow}
\usepackage{tabularx}
\usepackage{enumerate}
\usepackage{multicol}
\usepackage{hyperref}
\usepackage{subfig}
\usepackage{caption}

\def\bb0{{\mathbb{0}}}

\def\bb{{\mathbf{b}}}

\def\b0{{\mathbf{0}}}

\def\sf0{{\mathsf{0}}}

\def\rm0{{\mathrm{0}}}

\acrodef{CSI}[CSI]{channel state information}
\acrodef{CSIT}[CSIT]{channel state information at the transmitter}
\acrodef{CSIR}[CSIR]{channel state information at the receiver}
\acrodef{MIMO}[MIMO]{multiple-input multiple-output}
\acrodef{SISO}[SISO]{single-input single-output}
\acrodef{MISO}[MISO]{multiple-input single-output}
\acrodef{SIMO}[SIMO]{single-input multiple-output}
\acrodef{ADCs}[ADCs]{analog-to-digital convertors}
\acrodef{SNR}[SNR]{signal-to-noise ratio}
\acrodef{AWGN}[AWGN]{additive white Gaussian noise}
\acrodef{MRT}[MRT]{maximal ratio transmission}
\acrodef{DFT}[DFT]{Discrete Fourier Transform}
\acrodef{ULA}[ULA]{uniform linear array}
\acrodef{UPA}[UPA]{uniform planar array}
\acrodef{LS}[LS]{least squares}
\acrodef{ALMMSE}[ALMMSE]{approximate linear minimum mean squared error}
\acrodef{QIHT}[QIHT]{quantized iterative hard thresholding}
\acrodef{QIST}[QIST]{quantized iterative soft thresholding}
\acrodef{SVD}[SVD]{singular value decomposition}

\ifCLASSINFOpdf

\else
\fi
\begin{document}

\title{ A Model-Driven Deep Learning Network for \\ MIMO Detection}

\author{
\IEEEauthorblockN{Hengtao He\IEEEauthorrefmark{1},  Chao-Kai Wen\IEEEauthorrefmark{2}, Shi Jin\IEEEauthorrefmark{1}, Geoffrey~Ye~Li\IEEEauthorrefmark{3}}
\IEEEauthorblockA{\IEEEauthorrefmark{1}National Mobile Communications Research Laboratory, Southeast University\\ Nanjing 210096, P. R. China, E-mail: \{hehengtao, jinshi\}@seu.edu.cn}
\IEEEauthorblockA{\IEEEauthorrefmark{2}Institute of Communications Engineering, National Sun Yat-sen University\\ Kaohsiung 804, Taiwan, E-mail: chaokai.wen@mail.nsysu.edu.tw}
\IEEEauthorblockA{\IEEEauthorrefmark{3}School of Electrical and Computer Engineering,
Georgia Institute of Technology\\ Atlanta, GA 30332 USA, E-mail:
liye@ece.gatech.edu.}
}
%

%

\maketitle

\begin{abstract}

In this paper, we propose a model-driven deep learning network for multiple-input multiple-output (MIMO) detection. The structure of the network is specially designed by unfolding the iterative algorithm. Some trainable parameters are optimized through deep learning techniques to improve the detection performance. Since the number of trainable variables of the network is equal to that of the layers, the network can be easily trained within a very short time. Furthermore, the network can handle time-varying channel with only a single training. Numerical results show that the proposed approach can improve the performance of the iterative algorithm significantly under Rayleigh and correlated MIMO channels.


\end{abstract}


%
\IEEEpeerreviewmaketitle

\section{Introduction}
Multiple-Input Multiple-Output (MIMO) system has become mainstream technology
in many modern wireless communication standards, as it can increase the spectral efficiency and link reliability \cite{MIMO_capacity}. Efficient MIMO detection algorithms in terms of performance
and complexity play important roles in receiver design, and have arouse a series of
research \cite{SD,AMPJSTSP,EPdetector}. Maximum likelihood (ML) detection algorithm can achieve the optimal performance. However, it requires an exhaustive search, and its complexity increases exponentially with the number of decision variables, which prevents its deployment in practical systems. In order to reduce the complexity, sphere decoding has been proposed \cite{SD} by limiting the
search space. The performance of sphere decoding is quite
close to that of ML detector, but it still requires high complexity as the search space cannot be set too small.
Some suboptimal linear detectors, such as zero-forcing (ZF) and minimum mean-squared error (MMSE) detectors, are proposed. But they require relatively low complexity are far from optimal performance.

 Recently, iterative detectors based on approximate message passing (AMP) \cite{Donohol_AMP} and expectation propagation (EP) \cite{EP} have been proposed for  MIMO detection. The AMP-based detector \cite{AMPJSTSP} is simple and easy to be implemented in practice. It performs well when the elements of the MIMO channel matrix are with zero mean, independent identically distributed (i.i.d.) sub-Gaussian. The EP-based detector \cite{EPdetector} can achieve Bayes-optimal performance when the channel matrix is unitarily invariant, but it is with higher complexity than the AMP-based detector.

   Owing to strong learning ability from data, deep learning has been recently introduced to wireless physical layer \cite{DL2017wang}, such as millimeter wave channel estimation \cite{DL2018HE}, channel state information (CSI) feedback \cite{DL2018Wen}, and data detection \cite{DL2OFDM,DeepMIMO,Improving} , and achieved excellent performance. In particular, the detection network (DetNet) in \cite{DeepMIMO} unfolds the iterations of a projected gradient descent algorithm. The network architecture is designed from the iterative algorithm with trainable variables. The learning uses the result from the existing algorithm as an initial starting point and optimizes the variables based on back-propagation algorithms. Similar model-driven deep learning viewpoint has been successfully applied to image reconstruction \cite{ADMM_Net} and sparse signal recovery \cite{AMP_Net,TISTA} recently, and achieves better performance than iterative algorithms.

   This paper develops a model-driven deep learning network for  MIMO detection. 
   The structure of this network is obtained by adding some adjustable parameters to the existed iterative detectors and the performance can be improved by deep learning. Our network uses different linear estimators from \cite{TISTA} and more trainable parameters, which renders the network more flexible.
   Different from these works in \cite{AMP_Net,ICML}, which are only designed for a fixed channel, our network can perform MIMO detection under time-varying channels. Since few adjustable parameters are required to be optimized, the network can be easily trained within a very short time.
   Furthermore, this network  is very easy to achieve soft decision, which is more suitable for modern wireless communications system.
   The network is named OAMP-Net as it incorporates deep learning into the orthogonal AMP (OAMP) algorithm.
   The simulation results show that the OAMP-Net outperforms the OAMP algorithm significantly by learning the optimal parameters from a large number of data.

\emph{Notations}---For any matrix $\mathbf{A}$, $\mathbf{A}^{T}$ and ${ \mathrm{tr}}(\mathbf{A})$ denote  the transpose and the trace of $\mathbf{A}$, respectively. In addition, $\mathbf{I}$ is the identity matrix, $\mathbf{0}$ is the zero matrix, and $\mathbf{1}_n$ is the $n$-dimensional all-ones vector. A proper complex Gaussian with mean $\boldsymbol{\mu}$ and covariance $\boldsymbol{\Omega}$ can be described by the probability density function:
\begin{equation*}
  \mathcal{N}_{\mathbb{C}}(\mathbf{z};\boldsymbol{\mu},\boldsymbol{\Omega})=\frac{1}{\mathrm{det}(\pi \boldsymbol{\Omega})}
  e^{-(\mathbf{z}-\boldsymbol{\mu})^{H}\boldsymbol{\Omega}^{-1}(\mathbf{z}-\boldsymbol{\mu})},
\end{equation*}
and for a real Gaussian distribution
\begin{equation*}
\mathcal{N}(x;\mu, \sigma^{2})\triangleq\frac{1}{\sqrt{2\pi\sigma^{2}}}e^{-\frac{(x-\mu)^{2}}{2\sigma^{2}}}.
\end{equation*}

     The remaining part of this paper is organized as follows. Section \ref{Problem}
formulates the MIMO detection problem and introduces the OAMP algorithm. Next, the OAMP-Net are provided in Section \ref{OAMP-Net}. Then, Numerical results are presented in Section \ref{Simulation}. Finally, Section \ref{con} concludes
the paper.

\section{Problem Description and Algorithm Review}\label{Problem}
In this section, we consider an MIMO system with $N$ transmitting and $M$ receiving antennas. We formulate the MIMO detection problem by adopting the Bayesian inference. We also review the OAMP algorithm to apply it in the MIMO detection.
\subsection{MIMO detection}
 The transmitted symbol vector $\bar{\mathbf{x}}\in\mathbb{C}^{M\times1}$, each element drawn from the $P$-QAM constellation, is transmitted over a Rayleigh fading channel $\bar{\mathbf{H}}$, in which each element drawn from i.i.d. complex Gaussian distribution. The received signal $\bar{\mathbf{y}}\in\mathbb{C}^{N\times1}$ is given by
\begin{equation}\label{eq1}
  \bar{\mathbf{y}}=\bar{\mathbf{H}}\bar{\mathbf{x}}+\bar{\mathbf{n}},
\end{equation}
where $\bar{\mathbf{n} }\sim \mathcal{N}_{\mathbb{C}}(0,\sigma^{2}\mathbf{I}_{M})$ is the additive white Gaussian noise (AWGN).

As deep learning is always performed in the real-valued domain, we consider an equivalent real-valued representation which is obtained by considering the real $\mathfrak{R}(\cdot)$ and imaginary $\mathfrak{I}(\cdot)$ parts separately. Denote $\mathbf{x}=[\mathfrak{R}(\bar{\mathbf{x}})^{T}, \mathfrak{I}(\bar{\mathbf{x}})^{T}]^{T}$, $\mathbf{y}=[\mathfrak{R}(\bar{\mathbf{y}})^{T}, \mathfrak{I}(\bar{\mathbf{y}})^{T}]^{T}$, $\mathbf{n}=[\mathfrak{R}(\bar{\mathbf{n}})^{T}, \mathfrak{I}(\bar{\mathbf{n}})^{T}]^{T}$ and
\begin{equation}\label{eq2}
\mathbf{H}=\left[\begin{array}{c c c}
\mathfrak{R}(\bar{\mathbf{H}}) && -\mathfrak{I}(\bar{\mathbf{H}}) \\
\mathfrak{I}(\bar{\mathbf{H}}) && \mathfrak{R}(\bar{\mathbf{H}})
\end{array}\right].
\end{equation}
The system model can be rewritten in terms of real vectors and matrix as follows:
\begin{equation}\label{eq3}
\mathbf{y}=\mathbf{H}\mathbf{x}+\mathbf{n}.
\end{equation}

  We adopt the Bayesian inference to recover the signals $\mathbf{x}$ from the received signal $\mathbf{y}$. Based on the Bayes theorem, the posterior probability is given by
  \begin{equation}\label{eq4}
    \mathcal{P}(\mathbf{x}|\mathbf{y},\mathbf{H})=\frac{\mathcal{P}(\mathbf{y}|\mathbf{x},\mathbf{H})\mathcal{P}(\mathbf{x})}{\mathcal{P}(\mathbf{y})}
 = \frac{\mathcal{P}(\mathbf{y}|\mathbf{x},\mathbf{H})\mathcal{P}(\mathbf{x})}{\int \mathcal{P}(\mathbf{y}|\mathbf{x},\mathbf{H})\mathcal{P}(\mathbf{x})d\mathbf{x}}.
  \end{equation}
Given the posterior probability $\mathcal{P}(\mathbf{x}|\mathbf{y},\mathbf{H})$, the Bayesian MMSE estimate is obtained by
\begin{equation}\label{MMSE_estimate}
\hat{\mathbf{x}}=\int \mathbf{x} \mathcal{P}(\mathbf{x}|\mathbf{y},\mathbf{H})d\mathbf{x}.
\end{equation}

However, the Bayesian MMSE estimator is not computationally tractable because the marginal posterior probability in (\ref{MMSE_estimate}) involves a high-dimensional integral. In recent study \cite{OAMP}, the OAMP algorithm has been proposed as an iterative method to recover signal $\mathbf{x}$. We will show the OAMP-based detector in the following subsection.
\subsection{OAMP-based detector}
An OAMP algorithm has been proposed to solve sparse linear inverse problems in compressed sensing \cite{OAMP}. The principle of the algorithm is to decouple the posterior probability $\mathcal{P}(\mathbf{x}|\mathbf{y},\mathbf{H})$ into a series of
$\mathcal{P}(x_{i}|\mathbf{y},\mathbf{H}) (i=1,2,\ldots,2N)$ by an iterative way.  The OAMP-based detector is summarized in Algorithm 1.
\begin{algorithm}\label{algGE}
\caption{OAMP algorithm for MIMO detection} 
\hspace*{0.02in} {\bf Input:} 
Received signal $\mathbf{y}$, channel matrix $\mathbf{H}$, noise level $\sigma^{2}$. \\
\hspace*{0.02in} {\bf Output:} 
Recovered signal $\mathbf{x}_{t}$.\\
\hspace*{0.02in} {\bf Initialize:}
$\tau_{t} \leftarrow 1$, $\mathbf{x}_{t}\leftarrow \mathbf{0}$
\begin{equation}\label{eqr}
  \mathbf{r}_{t}=\hat{\mathbf{x}}_{t}+\mathbf{W}_{t}(\mathbf{y}-\mathbf{H}\hat{\mathbf{x}}_{t}),
\end{equation}
\begin{equation}\label{eqs}
  \hat{\mathbf{x}}_{t+1}=\mathbb{E}\left\{\mathbf{x}|\mathbf{\mathbf{r}}_{t},\tau_{t}\right\},
\end{equation}
\begin{equation}\label{eqv}
  v_{t}^{2}=\frac{\|\mathbf{y}-\mathbf{H}\hat{\mathbf{x}_{t}}\|_{2}^{2}-M\sigma^{2}}{\mathrm{tr}(\mathbf{H}^{T}\mathbf{H})},
\end{equation}

\begin{equation}\label{eqt}
  \tau^{2}_{t}=\frac{1}{2N}\mathrm{tr}(\mathbf{B}_{t}\mathbf{B}_{t}^{T})v_{t}^{2}+\frac{1}{4N}\mathrm{tr}(\mathbf{W}_{t}\mathbf{W}_{t}^{T})\sigma^{2}.
\end{equation}

\end{algorithm}

In Algorithm 1, the matrix $\mathbf{W}_{t}$ could be the transpose of $\mathbf{H}$, the pseudo inverse of $\mathbf{H}$, or the linear MMSE matrix. But the optimal one \cite{OAMP}  is given by
\begin{equation}\label{UMMSE}
  \mathbf{W}_{t}=\frac{2N}{\mathrm{tr}(\hat{\mathbf{W}_{t}}\mathbf{H})}\hat{\mathbf{W}_{t}},
\end{equation}
where $\hat{\mathbf{W}_{t}}$ is the linear MMSE matrix,
\begin{equation}\label{BMMSE}
 \hat{\mathbf{W}_{t}} = v_{t}^2\mathbf{H}^{T}(v_{t}^2\mathbf{H}\mathbf{H}^{T}+\frac{\sigma^{2}}{2}\mathbf{I})^{-1}.
\end{equation}
The matrix $\mathbf{W}_{t}$ is called de-correlated if $\mathrm{tr}(\mathbf{I}-\mathbf{W}_{t}\mathbf{H})=0$. In this case, the entries of  $\mathbf{h}_{t}=\mathbf{r}_{t}-\mathbf{x}$ are uncorrelated with those of $\mathbf{x}$ and mutually uncorrelated with zero-mean and identical variances.
The matrix $\mathbf{B}_{t}$ in the algorithm is given by $\mathbf{B}_{t}=\mathbf{I}-\mathbf{W}_{t}\mathbf{H}$.

   The posterior mean estimator in (\ref{eqs}) is with respect to the equivalent AWGN channel
  \begin{equation}\label{eqAWGN}
    \mathbf{r}_{t}=\mathbf{x}+ \mathbf{w}_{t},
  \end{equation}
  where $\mathbf{w}_{t}\sim \mathcal{N}(\mathbf{x};\mathbf{0},\tau^{2}_{t}\mathbf{I})$. It has been  indicated in \cite{OAMP} that the posterior mean estimator in $(\ref{eqs})$, which is usually non-linear and the linear estimator in (\ref{UMMSE})  are statistically orthogonal when the sensing matrix is unitarily invariant. The OAMP algorithm is Bayes-optimal, as it decouples the linear mixing model (\ref{eq3}) into $N$ parallel AWGN channel in (\ref{eqAWGN}) and uses (\ref{eqs}) to obtain the Bayesian MMSE estimate of $\mathbf{x}$.

  From (\ref{eqs}) and (\ref{eqAWGN}), we observe that $\mathbf{r}_{t}$ and $\tau^{2}_{t}$ are the prior mean and  variance  that influence the accuracy of  $\hat{\mathbf{x}}_{t+1}$. We will use a deep learning approach to provide appropriate step size to update $\mathbf{r}_{t}$ and $\tau^{2}_{t}$ and learn optimal variables from a large number of data.
\begin{figure}
  \centering
  \includegraphics[width=9cm]{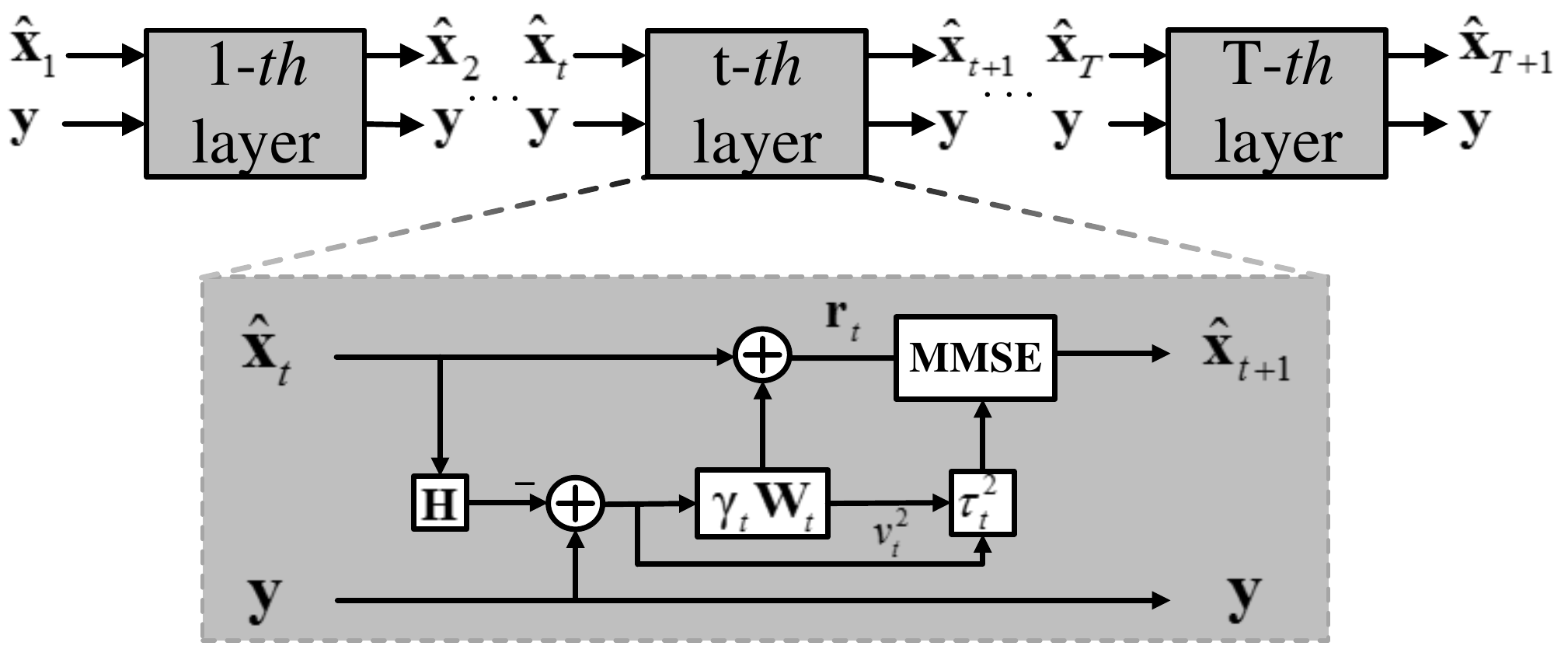}
  \caption{.~~The structure of the OAMP-Net network.}\label{fig1}
\end{figure}
\section{OAMP-Net}\label{OAMP-Net}

In this section, we present the OAMP-Net for MIMO detection. After discussing the function of each module and trainable variables, we analyze the computation complexity of the network.
\subsection{OAMP-Net architecture}
The structure of  the OAMP-Net is illustrated in Fig. 1, which is a revised version of algorithm by adding learnable scalar variables $\gamma_{t}$ and $\theta_{t}$. The network consists of $T$ cascade layers, and each layer has the same structure that contains the MMSE denoiser, error variance $\tau^{2}_{t}$, and tied weights. The input of the OAMP-Net are the received signal $\mathbf{y}$ and the initial value $\hat{\mathbf{x}}_{1}=\mathbf{0}$, and the output is the final estimate $\hat{\mathbf{x}}_{T+1}$ of signal $\mathbf{x}$.

For the $t$-th layer of the OAMP-Net, the input are the estimated signal $\hat{\mathbf{x}}_{t}$ from the $(t-1)$-th layer and the received signal $\mathbf{y}$, and data detection is performed as follows
\begin{equation}\label{eqlr}
  \mathbf{r}_{t}=\hat{\mathbf{x}}_{t}+\gamma_{t}\mathbf{W}_{t}(\mathbf{y}-\mathbf{H}\hat{\mathbf{x}}_{t}),
\end{equation}
\begin{equation}\label{eqls}
  \hat{\mathbf{x}}_{t+1}=\mathbb{E}\left\{\mathbf{x}|\mathbf{r}_{t},\tau_{t}\right\},
\end{equation}
\begin{equation}\label{eqlv}
  v_{t}^{2}=\frac{\|\mathbf{y}-\mathbf{H}\hat{\mathbf{x}_{t}}\|_{2}^{2}-M\sigma^{2}}{\mathrm{tr}(\mathbf{H}^{T}\mathbf{H})},
\end{equation}
\begin{equation}\label{eqlt}
  \tau^{2}_{t}=\frac{1}{2N}\mathrm{tr}(\mathbf{C}_{t}\mathbf{C}_{t}^{T})v_{t}^{2}+\frac{\theta_{t}^{2}\sigma^{2}}{4N}\mathrm{tr}(\mathbf{W}_{t}\mathbf{W}_{t}^{T}),
\end{equation}
where the error variance $\tau^{2}_{t}$ contains the contributions of error variance $v_{t}^{2}$ and trainable variable $\theta_{t}$, and $\mathbf{C}_{t}=\mathbf{I}-\theta_{t}\mathbf{W}_{t}\mathbf{H}$. $\mathbf{r}_{t}$  can be considered as the noisy observation from the equivalent AWGN channel (\ref{eqAWGN}). The scalar variables $(\gamma_{t},\theta_{t})$ are learnable variables that are optimized in the training process, and $v_{t}^{2}$ and $\tau^{2}_{t}$ are the error variances that describe the true error variances

\begin{equation}\label{eqvbar}
\bar{v}_{t}^{2}=\frac{\mathbb{E}[\|\mathbf{q}_{t}\|^{2}_{2}]}{2N}, \, \bar{\tau}_{t}^{2}=\frac{\mathbb{E}[\|\mathbf{h}_{t}\|^{2}_{2}]}{2N},
\end{equation}
where $\mathbf{q}_{t}=\hat{\mathbf{x}}_{t}-\mathbf{x}$. In OAMP-Net, we consider the matrix $\mathbf{W}_{t}$ as the optimal linear MMSE matrix in (\ref{UMMSE}).
The $\mathbb{E}\left\{\mathbf{x}|\mathbf{r}_{t},\tau_{t}\right\}$ is the MMSE denoiser, which is chosen according to the prior distribution of the original signal $\mathbf{x}$. If the transmitted symbol $\mathbf{x}$ is from the real alphabet set $\mathcal{S}=\{s_{1},s_{2},\ldots,s_{\sqrt{P}}\}$, corresponding posterior mean estimator for each element of $\hat{\mathbf{x}}$ is given by
\begin{equation}\label{eqE}
\mathbb{E}\left\{x_{i}|r_{i},\tau_{t}\right\}=\frac{\sum_{s_{i}}s_{i}\mathcal{N}(s_{i};r_{i}, \tau^{2}_{t})p(s_{i})}{\sum_{s_{i}}\mathcal{N}(s_{i};r_{i}, \tau^{2}_{t})p(s_{i})}.
\end{equation}

The only difference between the OAMP algorithm and OAMP-Net is the learnable variables $(\gamma_{t},\theta_{t} )$, which play  important roles in the network. OAMP algorithm assumes that $\gamma_{t}=\theta_{t}=1$ in order to ensure the orthogonality between the $\mathbf{q}_{t}$ and $\mathbf{h}_{t}$. Nevertheless, the orthogonality is guaranteed only when $\mathbf{H}$ is unitarily-invariant matrix. Because the function of learnable variables $(\gamma_{t},\theta_{t} )$ in (\ref{eqlr}) is providing appropriate step sizes for the update of mean and variance in the MMSE denoiser, we are intended to obtain optimal variables $(\gamma_{t},\theta_{t} )$ by deep learning. Furthermore,  when the matrix $\mathbf{W}_{t}$ is the pseudo inverse of $\mathbf{H}$ and $\gamma_{t}=\theta_{t}$ , the OAMP-Net is simplified to TISTA network \cite{TISTA}.

%
%

These error variance estimators in (\ref{eqlv}) and (\ref{eqlt})   play  important roles in providing appropriate variance estimates required for the MMSE denoiser. We only provide the final expressions for the error variances  because of space limitation. They can be derived by following two assumptions on the residual errors vector in \cite{OAMP}. The first assumption is that $\mathbf{h}_{t}$ consists of i.i.d. zero-mean Gaussian entries independent of $\mathbf{x}$.
Based on this assumption, each entry of the output from the linear estimator (\ref{eqlr}) can be regarded as an observation obtained from a virtual AWGN channel with the equivalent noise variance $\tau^{2}_{t}$.
Another assumption is that $\mathbf{q}_{t}$ consists of zero-mean i.i.d. entries independent of $\mathbf{H}$ and $\mathbf{n}$, which means that
\begin{equation}\label{eqA2}
\mathbb{E}[(\hat{\mathbf{x}}_{t}-\mathbf{x})^{T}\mathbf{H}^{T}\mathbf{n}]=0.
\end{equation}

If the calculation result in (\ref{eqlv}) is negative, we substitute $v_{t}^{2}$ by $\mathrm{max}(v_{t}^{2},\epsilon)$  for a small positive constant $\varepsilon$.
\subsection{Computation complexity}

The computation complexity required for the OAMP-Net per iteration is $O(N^3)$, similar to that of the OAMP algorithm. The computation complexity is dominant by the matrix inverse in $(\ref{UMMSE})$ in each iteration.

From the Fig.1,
the total number of trainable variables is equal to $2T$, since each layer of the OAMP-Net contains
only two adjustable variables $(\gamma_{t},\theta_{t} )$.
Furthermore, the number of trainable variables of the OAMP-Net is independent of the
number of antennas $N$ and $M$, and only determined by the number of layers $T$.
This is an advantageous feature for large-scale problems, such as high-dimensional MIMO detection.
With only few trainable variables, the stability and speed of convergence can be improved in the training process.

\section{Simulation Results}\label{Simulation}
In this section, we provide simulated  results of the OAMP-Net for MIMO detection. The signal-to-noise (SNR) of the system, defined as
\begin{equation}\label{eqsnr}
  \mathrm{SNR}=\frac{\mathbb{E}\|\mathbf{H}\mathbf{x}\|^{2}_{2}}{\mathbb{E}\|\mathbf{n}\|^{2}_{2}},
\end{equation}
is used to measure the noise level.
\subsection{Implementation details}
In our simulation, the OAMP-Net is implemented in Tensorflow. 
The number of layers $T$ is set to $10$. The training data consists of a number of randomly generated pairs $(\mathbf{x},\mathbf{y})$. The data $\mathbf{x}$ is generated from QPSK modulation symbol. We train the network with $10,000$ epochs. At each epoch, the training and validation sets contain $5,000$, $1,000$ samples, respectively.
For test sets, we generate the test data to test the network until the number of bit errors exceed $1,000$. The OAMP-Net is trained using the stochastic gradient descent method and Adam optimizer. The learning rate is set to be $0.001$. The batch size is set to $1000$. Furthermore, we set $\varepsilon=10^{-9}$ to avoid stability problem.
In our experiment settings, we choose the $L_{2}$ loss as the cost function.

\subsection{Rayleigh MIMO channel}
In this section, we consider a point-to-point MIMO system with $M$ transmitting and $N$ receiving antennas, where $M=N$.
The channel $\mathbf{H}$ is time-varying and each element drawn from $\mathbf{H}\sim\mathcal{N}_{\mathbb{C}}(0,1/M)$, i.e., each component of the channel matrix $\mathbf{H}$ obeys a zero-mean Gaussian distribution with variance $1/M$.

\begin{figure}
  \centering
  \includegraphics[width=2.85in]{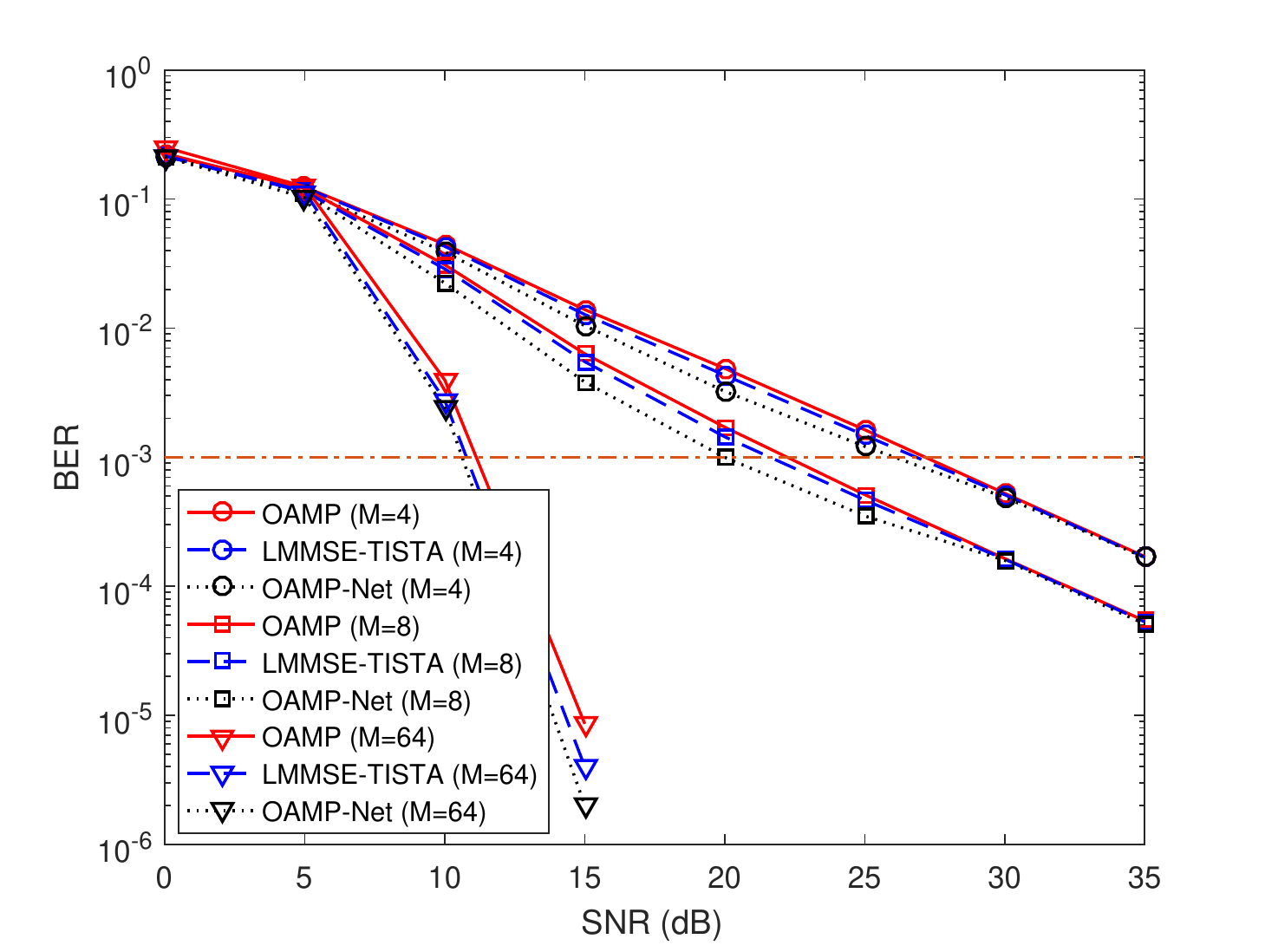}
  \caption{.~~BER versus SNR for OAMP algorithm and OAMP-Net with Rayleigh MIMO channel.}\label{fig2}
\end{figure}

Fig. \ref{fig2} compares the average bit-error rate (BER) performance of the OAMP algorithm, the LMMSE-TISTA network and OAMP-Net \cite{EP}.
The LMMSE-TISTA network is a variant of TISTA network \cite{TISTA} 
by replaceing the pseudo inverse matrix with the linear MMSE estimator. From the figure, the OAMP-Net outperforms the OAMP algorithm in all setting, which demonstrates the deep learning can improve the OAMP-based detector. The reason for the performance improvement is that the fixed parameters $(\gamma_{t},\theta_{t})$ in the OAMP algorithm is trainable in the OAMP-Net in each layer,  which renders the network more flexible. 
Specifically, if we target the SNR for BER=$10^{-3}$, BER performance improves about $1.37$ dB by deep learning when $M=N=4$. By contrast, the gains are approximately $2.97$ dB and $0.82$ dB when $M=N=8$ and $M=N=64$, respectively. 
\subsection{Correlated MIMO channel}
We consider the correlated MIMO channel in this section, which can be described by the Kronecker model,
\begin{equation}\label{eqcor}
  \mathbf{H}=\mathbf{R}_R^{1/2}\mathbf{A}\mathbf{R}_T^{1/2},
\end{equation}
where $\mathbf{R}_{R}$ and $\mathbf{R}_{T}$ are the spatial correlation matrix at the receiver and the transmitter, which are generated according to the exponential correlation model \cite{corMIMO} with correlation coefficient $\rho=0.5$, and $\mathbf{A}$ is the Rayleigh fading channel matrix. We consider the small size MIMO system with $M=N=4$.

\begin{figure}
  \centering
  \includegraphics[width=2.85in]{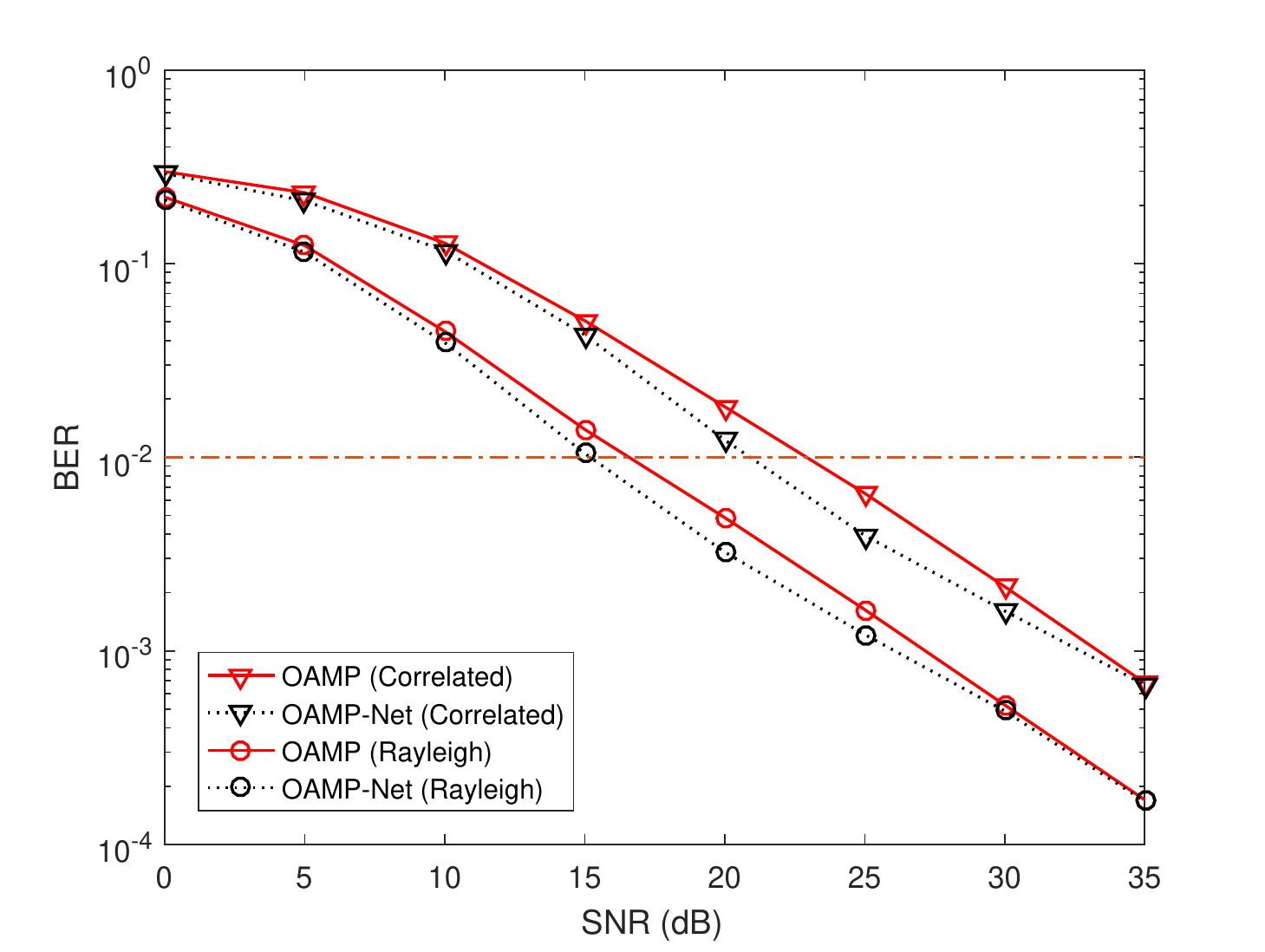}
  \caption{.~~BER versus SNR for OAMP algorithm and OAMP-Net with correlated MIMO channel.}\label{fig3}
\end{figure}

 Fig. \ref{fig3} illustrates the BER performance of the OAMP-Net and the OAMP algorithm under Rayleigh and correlated MIMO channels. In that case, all algorithms have performance degradation compared with the independent Rayleigh channel. For example, approximately $6.34$ dB and $6.05$ dB loss are caused due to the channel correlation for the OAMP-Net and OAMP algorithm, respectively, when we target the SNR for BER=$10^{-2}$. Furthermore,  the OAMP-Net can obtain more performance gain even if the channel has correlation. Compared with $1.86$ dB gain obtained in the independent Rayleigh channel, the OAMP-Net can obtain more than $2.15$ dB performance improvement under the correlated MIMO channel.

\section{Conclusion}\label{con}
We have developed a novel model-driven deep learning network for  MIMO detection. This network inherits the superiority of the Bayes-optimal signal recovery algorithm and deep learning techniques, and thus presents excellent performance. The network is easy and fast to train because only few adjustable parameters are required to be optimized. Furthermore, this network can handle the time-varying channel. 
Simulation results demonstrate that the network outperforms the OAMP algorithm significantly under the independent Rayleigh and correlated MIMO channel, which shows deep learning can improve the iterative algorithm by optimizing some parameters.

\section*{Acknowledgment}
The work was supported by the National Science
Foundation (NSFC) for Distinguished Young Scholars of China with Grant
61625106 and the NSFC with Grant 61531011.

\end{document}